\documentstyle[amssymb,12pt]{article}

\if@twoside
\else
\fi
\advance\textheight by \topskip
\leftmargin\leftmargini
\labelwidth\leftmargini\advance\labelwidth-\labelsep

\catcode`\@=11
\def\Let@{\relax\iffalse{\fi\let\\=\cr\iffalse}\fi}
\def\vspace@{\def\vspace##1{\crcr\noalign{\vskip##1\relax}}}
\def\multilimits@{\bgroup\vspace@\Let@
 \baselineskip\fontdimen10 \scriptfont\tw@
 \advance\baselineskip\fontdimen12 \scriptfont\tw@
 \lineskip\thr@@\fontdimen8 \scriptfont\thr@@
 \lineskiplimit\lineskip
 \vbox\bgroup\ialign\bgroup\hfil$\m@th\scriptstyle{##}$\hfil\crcr}
\def\Sb{_\multilimits@}
\def\endSb{\crcr\egroup\egroup\egroup}
\def\Sp{^\multilimits@}

\long
\def\QQQ#1#2{}
\def\QTP#1{}
\long
\def\QQA#1#2{}

\def\EXPAND#1[#2]#3{}
\def\NOEXPAND#1[#2]#3{}

\def\LaTeXparent#1{}

\newdimen\ex@
\def\rightarrowfill@#1{$#1\m@th\mathord-\mkern-6mu\cleaders
 \hbox{$#1\mkern-2mu\mathord-\mkern-2mu$}\hfill
 \mkern-6mu\mathord\rightarrow$}%
\def\leftarrowfill@#1{$#1\m@th\mathord\leftarrow\mkern-6mu\cleaders
 \hbox{$#1\mkern-2mu\mathord-\mkern-2mu$}\hfill\mkern-6mu\mathord-$}%
\def\leftrightarrowfill@#1{$#1\m@th\mathord\leftarrow
\mkern-6mu\cleaders
 \hbox{$#1\mkern-2mu\mathord-\mkern-2mu$}\hfill
 \mkern-6mu\mathord\rightarrow$}%
\def\overrightarrow{\mathpalette\overrightarrow@}%
\def\overrightarrow@#1#2{\vbox{\ialign{##\crcr\rightarrowfill@#1\crcr
 \noalign{\kern-\ex@\nointerlineskip}$\m@th\hfil#1#2\hfil$\crcr}}}%

\def\overleftarrow{\mathpalette\overleftarrow@}%
\def\overleftarrow@#1#2{\vbox{\ialign{##\crcr\leftarrowfill@#1\crcr
 \noalign{\kern-\ex@\nointerlineskip}$\m@th\hfil#1#2\hfil$\crcr}}}%
\def\overleftrightarrow{\mathpalette\overleftrightarrow@}%
\def\overleftrightarrow@#1#2{\vbox{\ialign{##\crcr
   \leftrightarrowfill@#1\crcr
 \noalign{\kern-\ex@\nointerlineskip}$\m@th\hfil#1#2\hfil$\crcr}}}%
\def\underrightarrow{\mathpalette\underrightarrow@}%
\def\underrightarrow@#1#2{\vtop{\ialign{##\crcr$\m@th\hfil#1#2\hfil
  $\crcr\noalign{\nointerlineskip}\rightarrowfill@#1\crcr}}}%

\def\underleftarrow{\mathpalette\underleftarrow@}%
\def\underleftarrow@#1#2{\vtop{\ialign{##\crcr$\m@th\hfil#1#2\hfil
  $\crcr\noalign{\nointerlineskip}\leftarrowfill@#1\crcr}}}%
\def\underleftrightarrow{\mathpalette\underleftrightarrow@}%
\def\underleftrightarrow@#1#2{\vtop{\ialign{##\crcr$\m@th
  \hfil#1#2\hfil$\crcr
 \noalign{\nointerlineskip}\leftrightarrowfill@#1\crcr}}}%

\QQQ{Language}{
American English
}

\begin{document}

\bigskip

\begin{center}
{\LARGE {Duffin-Kemmer}-Petiau equation in Riemannian space-times}
\end{center}

\vspace{1cm}

\begin{center}
{\large J. T. Lunardi}\footnote{%
On leave from Departamento de Matem\'{a}tica e Estat\'{\i}stica. Setor de
Ci\^{e}ncias Exatas e Naturais. Universidade Estadual de Ponta Grossa. Ponta
Grossa, PR - Brazil}\footnote{%
E-mail: lunardi@ift.unesp.br}{\large , B. M. Pimentel}\footnote{%
E-mail: pimentel@ift.unesp.br} {\large and R. G. Teixeira}\footnote{%
E-mail: randall@ift.unesp.br}\\[0pt]

\vskip 0.5cm

Instituto de F\'{\i}sica Te\'{o}rica\\[0pt]
Universidade Estadual Paulista\\[0pt]
Rua Pamplona 145\\[0pt]
01405-900 - S\~{a}o Paulo, S.P.\\[0pt]
Brazil\\[0pt]
\end{center}

\vskip 1.5cm

\begin{center}
\begin{minipage}{14.5cm}
\centerline{\bf Abstract}
{In this work we analyze the generalization of Duffin-Kemmer-Petiau equation to the 
case of Riemannian space-times and show that the usual results for Klein-Gordon 
and Proca equations in Riemannian space-times can be fully recovered when one 
selects, respectively, the spin 0 and 1 sectors of Duffin-Kemmer-Petiau theory.}
\end{minipage}
\end{center}

\vskip1.5cm

PACS: 11.10.-z, 03.65.Pm

\newpage

\section{Introduction}

The Duffin-Kemmer-Petiau (DKP) equation is a convenient relativistic wave
equation to describe spin 0 and 1 bosons with the advantage over standard
relativistic equations, like Klein-Gordon (KG) and Proca ones, of being of
first order in derivatives. As a matter of fact, this equation was developed
specifically to fulfill this characteristic, and provide an equation for
bosons similar to Dirac spin 1/2 equation.

The first one to propose what is now known as the $16\times 16$ DKP algebra
was Petiau \cite{Petiau}, a de Broglie's student, who took as starting point
the former's work on first order wave equations on $16\times 16$ matrices
that were products of different Dirac matrices spaces. Latter it was showed
that this algebra could be decomposed into 5, 10 and 1 degrees irreducible
representations, the latter one being trivial \cite{Geheniau}. Anyway,
Petiau's work remained unknown to the majority of scientific community so
that Kemmer, working independently, wrote Proca's equation as a set of
coupled first order equations as well as the equivalent spin 0 case \cite
{Kemmer 1}. Although these set of equations could be written in $10\times 10$
(spin 1) and $5\times 5$ (spin 0) matrix forms, it was not clear which
algebraic relations these matrices should obey. Based on this work, Duffin
was able to put the equations sets in a first order $\beta $ matrix
formulation presenting 3 of the 4 commutation relations present in the DKP
algebra \cite{Duffin}. This result provided the motivation to Kemmer to
complete formalism and present the complete theory of a relativistic wave
equation for spin 0 and 1 bosons \cite{Kemmer 2}. These and other facts
related with the historical development of DKP theory, as well as a detailed
list of references on the subject, can be found in Krajcik and Nieto paper
on historical development of Bhabha first order relativistic equations \cite
{Krajcik}.

More recently there have been an increasing interest in DKP theory.
Specifically, it has been applied to QCD (large and short distances)\ by
Gribov \cite{Gribov}, to covariant Hamiltonian dinamics by Kanatchikov \cite
{Kanatchikov} and has been used in its generally relativistic version by
Red'kov to study, using a specific representation of the 10 degrees DKP
algebra, spin 1 particles in the abelian monopole field \cite{Red'kov}.

But, although DKP equation is known to be completely equivalent to KG and
Proca in the free field case, doubts arise, specially with respect to KG
equation, when minimal interaction with eletromagnetic field comes into
play. This equivalence has been proved recently at classical level \cite
{Nowakowski, Pimentel 1} and at classical and quantum levels too by Fainberg
\& Pimentel \cite{Pimentel 2}; but leaves open the question about whether
other processes or interactions could distinguish DKP from KG and Proca
equations. So, our intention in this work is to analyze the generalization
to Riemannian space-times of DKP equation and its equivalence to the
Riemannian versions of KG and Proca equations. This demonstration will be
performed by showing that one obtains the generalized KG and Proca equations
when the spin 0 and 1 sectors of the generalized DKP\ equation are selected.

In order to make this work more self contained we will dedicate some space
to basic results on DKP theory and to the construction of its Riemannian
generalization using the tetrad formalism. So, we start by presenting in
section 2 the basic results on DKP equation on Minkowski space-time. We
shall not enter in full details of the theory but simply quote the most
important results and properties, specially about the projectors of physical
spin 0 and 1 components of DKP field, necessary to the understanding of this
work. For further details we suggest the reader to original works \cite
{Petiau, Duffin, Kemmer 2} or classic textbooks \cite{Umezawa, Berestetskii}%
. In section 3 we will present some basic results on the tetrad formalism
and perform the passage from Minkowskian to Riemannian space-times. In
section 4 the equivalence between DKP equation and KG and Proca equations in
Riemannian space-times will be demonstrated using the projectors of physical
spin 0 and 1 sectors of DKP theory. Finally, we present our conclusions and
comments in section 5. Throughout this work we will adopt the signature $%
\left( +---\right) $ to the metric tensors as well as Einstein implicit
summation rule, except otherwise stated. Moreover, Latin letters will be
used when labelling indexes concerned to the Minkowski space-time or to the
Minkowskian manifolds tangent to the Riemannian manifold, while Greek
letters will label indexes referred to the Riemannian manifold. Both Latin
and Greek indexes will run from 0 to 3, except when we clearly state the
opposite.

\section{DKP equation in Minkowski space-time}

The DKP equation is given by 
\begin{equation}
\left( i\beta ^{a}\partial _{a}-m\right) \psi =0,  \label{eq1}
\end{equation}
where the matrices $\beta ^{a}$ obey the algebraic relations 
\begin{equation}
\beta ^{a}\beta ^{b}\beta ^{c}+\beta ^{c}\beta ^{b}\beta ^{a}=\beta ^{a}\eta
^{bc}+\beta ^{c}\eta ^{ba},  \label{eq2}
\end{equation}
being $\eta ^{ab}$ the metric tensor of Minkowski space-time. This equation
is very similar to Dirac's equation but the algebraic properties of $\beta
^{a}$ matrices, which have no inverses, make it more difficult to deal with.
From the algebraic relation above we can obtain (no summation on repeated
indexes) 
\begin{equation}
\left( \beta ^{a}\right) ^{3}=\eta ^{aa}\beta ^{a},  \label{eq3}
\end{equation}
so that we can define the matrices 
\begin{equation}
\eta ^{a}=2\left( \beta ^{a}\right) ^{2}-\eta ^{aa}  \label{eq3b}
\end{equation}
that satisfy 
\begin{equation}
\left( \eta ^{a}\right) ^{2}=1,\ \eta ^{a}\eta ^{b}-\eta ^{b}\eta ^{a}=0
\label{eq3c}
\end{equation}
\begin{equation}
\eta ^{a}\beta ^{b}+\beta ^{b}\eta ^{a}=0\ \left( a\neq b\right) ,
\label{eq4}
\end{equation}
\begin{equation}
\eta ^{aa}\beta ^{a}=\eta ^{a}\beta ^{a}=\beta ^{a}\eta ^{a}.  \label{eq5}
\end{equation}

With these results we can write the Lagrangian density for DKP field as 
\begin{equation}
{\cal L}=\frac{i}{2}\overline{\psi }\beta ^{a}\overleftrightarrow{\partial }%
_{a}\psi -m\overline{\psi }\psi ,  \label{eq6}
\end{equation}
where $\overline{\psi }$ is defined\ as 
\begin{equation}
\overline{\psi }=\psi ^{\dagger }\eta ^{0}  \label{eq6b}
\end{equation}
From this Lagrangian we can obtain DKP equation through a variational
principle. Moreover, we will choose $\beta ^{0}$ to be hermitian and $\beta
^{i}$ $\left( i=1,2,3\right) $\ anti-hermitian so that the equation for $%
\overline{\psi }$ can also be easily obtained by applying hermitian
conjugation to equation (\ref{eq1}).

Besides that, under a Lorentz transformation $x^{\prime a}=\Lambda
^{a}{}_{b}x^{b}$ we have 
\begin{equation}
\psi \rightarrow \psi ^{\prime }=U\left( \Lambda \right) \psi ,  \label{eq7}
\end{equation}
\begin{equation}
U^{-1}\beta ^{a}U=\Lambda ^{a}{}_{b}\beta ^{b},  \label{eq8}
\end{equation}
which gives, for infinitesimal transformations $\Lambda ^{ab}=\eta
^{ab}+\omega ^{ab}$ $\left( \omega ^{ab}=-\omega ^{ba}\right) $ \cite
{Umezawa}, 
\begin{equation}
U=1+\frac{1}{2}\omega ^{ab}S_{ab},\ S_{ab}=\left[ \beta _{a},\beta _{b}%
\right] .  \label{eq9}
\end{equation}

If one uses two sets of Dirac matrices $\gamma ^{a}$ and $\gamma ^{\prime a}$
acting on different indexes of a 16 component $\psi $ wave function it can
be verified that the matrices 
\begin{equation}
\beta ^{a}=\frac{1}{2}\left( \gamma ^{a}I^{\prime }+I\gamma ^{\prime
a}\right)  \label{eq10}
\end{equation}
satisfy the algebraic relation (\ref{eq2}), but these matrices form a
reducible representation since it can be shown \cite{Kemmer 2, Umezawa} that
this algebra has 3 inequivalent irreducible representations: a trivial 1
degree $\left( \beta ^{a}=0\right) $ without physical significance; a 5
degree one, corresponding to a 5 component $\psi $ that describes a spin 0
boson; and a 10 degree one, corresponding to a 10 component $\psi $
describing a spin 1 boson.

\subsection{The spin 0 sector of DKP theory}

The spin 0 sector can be selected from a general representation of $\beta $
matrices through the operators 
\begin{equation}
P=-\left( \beta ^{0}\right) ^{2}\left( \beta ^{1}\right) ^{2}\left( \beta
^{2}\right) ^{2}\left( \beta ^{3}\right) ^{2},  \label{eq11}
\end{equation}
which satisfies $P^{2}=P$, and 
\begin{equation}
P^{a}=P\beta ^{a}.  \label{eq12}
\end{equation}

It can be shown \cite{Umezawa} that 
\begin{equation}
P^{a}\beta ^{b}=P\eta ^{ab},\ PS^{ab}=S^{ab}P=0,\ P^{a}S^{bc}=\left( \eta
^{ab}P^{c}-\eta ^{ac}P^{b}\right) ,  \label{eq13}
\end{equation}
and, as a consequence, under infinitesimal Lorentz transformations (\ref{eq9}%
) we have 
\begin{equation}
PU\psi =P\psi ,  \label{eq14}
\end{equation}
so that $P\psi $ transforms as a (pseudo)scalar. Similarly 
\begin{equation}
P^{a}U\psi =P^{a}\psi +\omega ^{a}{}_{b}P^{b}\psi ,  \label{eq15}
\end{equation}
showing that $P^{a}\psi $ transforms like a (pseudo)vector.

Applying these operators to DKP equation (\ref{eq1}) we have 
\begin{equation}
\partial _{a}\left( P^{a}\psi \right) =\frac{m}{i}P\psi ,  \label{eq16}
\end{equation}
and 
\begin{equation}
P^{b}\psi =\frac{i}{m}\partial ^{b}\left( P\psi \right) ,  \label{eq17}
\end{equation}
which combined provide 
\begin{equation}
\partial _{a}\partial ^{a}\left( P\psi \right) +m^{2}\left( P\psi \right)
=\square \left( P\psi \right) +m^{2}\left( P\psi \right) =0.  \label{eq18}
\end{equation}

These results show that all elements of the column matrix $P\psi $ are
scalar fields of mass $m$ obeying KG equation while the elements of $%
P_{a}\psi $ are $\frac{i}{m}$ times the derivative with respect to $x^{a}$
of the corresponding elements of $P\psi $. Moreover, we can choose a 5
degree irreducible representation of the $\beta ^{a}$ matrices in such a way
that 
\begin{equation}
P\psi =P\left( 
\begin{array}{c}
\psi _{0} \\ 
\vdots \\ 
\psi _{4}
\end{array}
\right) =\left( 
\begin{array}{c}
0_{4\times 1} \\ 
\psi _{4}
\end{array}
\right) ,\ P_{a}\psi =\left( 
\begin{array}{c}
0_{4\times 1} \\ 
\psi _{a}
\end{array}
\right) ,  \label{eq19}
\end{equation}
so that equation (\ref{eq17}) results in 
\begin{equation}
\psi _{a}=\frac{i}{m}\partial _{a}\psi _{4}  \label{eq19b}
\end{equation}
allowing us to make $\psi _{4}=\sqrt{m}\varphi $ and obtain 
\begin{equation}
\psi =\left( 
\begin{array}{c}
\frac{i}{\sqrt{m}}\partial _{a}\varphi \\ 
\sqrt{m}\varphi
\end{array}
\right) ,\ \square \varphi +m^{2}\varphi =0,  \label{eq20}
\end{equation}
making evident that the DKP equation describes a scalar particle.

\subsection{The spin 1 sector of DKP theory}

In order to select the spin 1 sector of DKP equation from a general
representation of $\beta $ matrices we can use the operators 
\begin{equation}
R^{a}=\left( \beta ^{1}\right) ^{2}\left( \beta ^{2}\right) ^{2}\left( \beta
^{3}\right) ^{2}\left[ \beta ^{a}\beta ^{0}-\eta ^{a0}\right] ,
\label{eq20-2}
\end{equation}
and 
\begin{equation}
R^{ab}=R^{a}\beta ^{b}.  \label{eq20-3}
\end{equation}

From the definitions it can be shown \cite{Umezawa} that these operators
have the following properties 
\begin{equation}
R^{ab}=-R^{ba},  \label{eq20-4}
\end{equation}
\begin{equation}
R^{a}\beta ^{b}\beta ^{c}=R^{ab}\beta ^{c}=\eta ^{bc}R^{a}-\eta ^{ac}R^{b},
\label{eq20-5}
\end{equation}
\begin{equation}
R^{a}S^{bc}=\eta ^{ab}R^{c}-\eta ^{ac}R^{b},\ S^{bc}R^{a}=0,  \label{eq20-6}
\end{equation}
and 
\begin{equation}
R^{ab}S^{cd}=\eta ^{bc}R^{ad}-\eta ^{ac}R^{bd}-\eta ^{bd}R^{ac}+\eta
^{ad}R^{bc}.  \label{eq20-7}
\end{equation}

Then, under infinitesimal Lorentz transformations (\ref{eq9}), we have 
\begin{equation}
R^{a}U\psi =R^{a}\psi +\omega ^{a}{}_{b}R^{b}\psi  \label{eq20-8}
\end{equation}
and 
\begin{equation}
R^{ab}U\psi =R^{ab}\psi +\omega ^{b}{}_{c}R^{ac}\psi +\omega
^{a}{}_{c}R^{cb}\psi ,  \label{eq20-9}
\end{equation}
showing that $R^{a}\psi $ transforms like a (pseudo)vector while $R^{ab}\psi 
$ transforms like a rank 2 (pseudo)tensor.

The application of these operators to DKP equation results in 
\begin{equation}
\partial _{b}\left( R^{ab}\psi \right) =\frac{m}{i}R^{a}\psi ,
\label{eq20-10}
\end{equation}
and 
\begin{equation}
R^{ab}\psi =-\frac{i}{m}U^{ab},  \label{eq20-11}
\end{equation}
where 
\begin{equation}
U^{ab}=\partial ^{a}R^{b}\psi -\partial ^{b}R^{a}\psi  \label{eq20-12}
\end{equation}
is the strength tensor of the massive vector field $R^{a}\psi $. Combined,
these results provide 
\begin{equation}
\partial _{b}\left( -\frac{i}{m}U^{ab}\right) =\frac{m}{i}R^{a}\psi
\label{eq20-13}
\end{equation}
\begin{equation}
\partial _{b}U^{ba}+m^{2}R^{a}\psi =0,  \label{eq20-14}
\end{equation}
or equivalently 
\begin{equation}
\left( \square +m^{2}\right) R^{a}\psi =0;\ \partial _{a}R^{a}\psi =0.
\label{eq20-15}
\end{equation}

So, all elements of the column matrix $R^{a}\psi $ are components vector
fields of mass $m$ obeying Proca equation; being the elements of $R^{ab}\psi 
$ equal to $\frac{-i}{m}$ times the field strength tensor of the vector
field of which the corresponding elements of $R^{a}\psi $ are components.
So, similarly to the spin 0 case, this procedure selects the spin 1 content
of DKP theory, making explicitly clear that it describes a massive vectorial
particle.

Similarly to the spin 0 sector of DKP theory, we can find a 10 degree
irreducible representation of the $\beta ^{a}$ matrices in such a way that

\begin{equation}
R_{a}\psi =R_{a}\left( 
\begin{array}{c}
\psi _{0} \\ 
\vdots \\ 
\psi _{9}
\end{array}
\right) =\left( 
\begin{array}{c}
\psi _{a} \\ 
0_{9\times 1}
\end{array}
\right) .  \label{eq20-16}
\end{equation}
\begin{equation}
R_{10}\psi =\left( 
\begin{array}{c}
\psi _{7} \\ 
0_{9\times 1}
\end{array}
\right) ,R_{20}\psi =\left( 
\begin{array}{c}
\psi _{8} \\ 
0_{9\times 1}
\end{array}
\right) ,R_{30}\psi =\left( 
\begin{array}{c}
\psi _{9} \\ 
0_{9\times 1}
\end{array}
\right) ,  \label{eq20-17}
\end{equation}
\begin{equation}
R_{12}\psi =\left( 
\begin{array}{c}
\psi _{6} \\ 
0_{9\times 1}
\end{array}
\right) ,R_{31}\psi =\left( 
\begin{array}{c}
\psi _{5} \\ 
0_{9\times 1}
\end{array}
\right) ,R_{23}\psi =\left( 
\begin{array}{c}
\psi _{4} \\ 
0_{9\times 1}
\end{array}
\right)  \label{eq20-18}
\end{equation}

Defining 
\begin{equation}
\psi _{a}=\sqrt{m}B_{a},  \label{eq20-19}
\end{equation}
where $B_{a}$ is a vector field, we get from equation (\ref{eq20-12}) that 
\begin{equation}
U_{ab}=\sqrt{m}\left( 
\begin{array}{c}
\partial _{a}B_{b}-\partial _{b}B_{a} \\ 
0_{9\times 1}
\end{array}
\right) =\sqrt{m}\left( 
\begin{array}{c}
K_{ab} \\ 
0_{9\times 1}
\end{array}
\right) ,  \label{eq20-20}
\end{equation}
where 
\begin{equation}
K_{ab}=\left( \partial _{a}B_{b}-\partial _{b}B_{a}\right) .  \label{eq20-21}
\end{equation}

Consequently equation (\ref{eq20-11}) will result in 
\begin{equation}
R_{ab}\psi =-\frac{i}{\sqrt{m}}\left( 
\begin{array}{c}
K_{ab} \\ 
0_{9\times 1}
\end{array}
\right) ,  \label{eq20-22}
\end{equation}
which, together with equations (\ref{eq20-17} and (\ref{eq20-18}),
determines the components $\psi _{4}$ to $\psi _{9}$ in $\psi $. So equation
(\ref{eq20-14}) can now be written as 
\begin{equation}
\partial _{a}\left( K^{ab}\right) +m^{2}\left( B^{b}\right) =0,
\label{eq20-23}
\end{equation}
making explicit that the DKP equation describes a massive vector field.

\section{Passage to Riemannian space-times}

Before constructing the DKP equation in Riemannian ${\cal R}^{4}$
space-times we will construct the tensor quantities on ${\cal R}^{4}$ using
the tensors defined on a Minkowski manifold tangent to each point of ${\cal R%
}^{4}$, and for this purpose we will use the standard formalism of
``tetrads''. Here we will just mention the necessary fundamental results and
point out the reader to reference \cite{Sabbata}.

A\ tetrad is constituted by a set of four vector fields $e_{\mu
}{}^{a}\left( x\right) $ that satisfy, at each point $x$ of ${\cal R}^{4}$,
the relations 
\begin{equation}
\eta ^{ab}=e_{\mu }{}^{a}\left( x\right) e_{\nu }{}^{b}\left( x\right)
g^{\mu \nu }\left( x\right) ,  \label{eq21}
\end{equation}
\begin{equation}
g_{\mu \nu }\left( x\right) =e_{\mu }{}^{a}\left( x\right) e_{\nu
}{}^{b}\left( x\right) \eta _{ab},  \label{eq22}
\end{equation}
and 
\begin{equation}
\eta _{ab}=e^{\mu }{}_{a}\left( x\right) e^{\nu }{}_{b}\left( x\right)
g_{\mu \nu }\left( x\right) ,  \label{eq23}
\end{equation}
\begin{equation}
g^{\mu \nu }\left( x\right) =e^{\mu }{}_{a}\left( x\right) e^{\nu
}{}_{b}\left( x\right) \eta ^{ab},  \label{eq24}
\end{equation}
where 
\begin{equation}
e^{\mu }{}_{a}\left( x\right) =g^{\mu \nu }\left( x\right) \eta _{ab}e_{\nu
}{}^{b}\left( x\right) ,  \label{eq25}
\end{equation}
the Latin indexes being raised and lowered by \ the Minkowski metric $\eta
^{ab}$ and the Greek ones by the metric $g^{\mu \nu }$ of the manifold $%
{\cal R}^{4}$.

The components $B^{ab}$ in ${\cal M}^{4}$ of a tensor $B^{\mu \nu }$ defined
on ${\cal R}^{4}$ are given by 
\begin{equation}
B^{ab}=e_{\mu }{}^{a}e_{\nu }{}^{b}B^{\mu \nu },\ B_{ab}=e^{\mu
}{}_{a}e^{\nu }{}_{b}B_{\mu \nu },  \label{eq26}
\end{equation}
or inversely 
\begin{equation}
B^{\mu \nu }=e^{\mu }{}_{a}e^{\nu }{}_{b}B^{ab},\ B_{\mu \nu }=e_{\mu
}{}^{a}e_{\nu }{}^{b}B_{ab},  \label{eq27}
\end{equation}
and it is easy to see that $A_{\mu }B^{\mu }=A_{a}B^{a}$.

The Lorentz covariant derivative $D_{\mu }$ is defined as 
\begin{equation}
D_{\mu }B^{a}=\partial _{\mu }B^{a}+\omega _{\mu }{}^{a}{}_{b}\left(
x\right) B^{b},  \label{eq28}
\end{equation}
such that $D_{\mu }B^{a}$ transforms like a vector under local Lorentz
transformations $x^{a}=\Lambda ^{a}{}_{b}\left( x\right) x^{b}$ on the $%
{\cal M}^{4}$ tangent manifold, where the connection $\omega _{\mu
}{}^{ab}\left( x\right) $ on ${\cal M}^{4}$ satisfies the transformation
rule 
\begin{equation}
\omega _{\mu }^{\prime }{}^{ab}=\Lambda ^{a}{}_{c}\omega _{\mu
}{}^{c}{}_{l}\left( \Lambda ^{-1}\right) ^{lb}-\left( \partial _{\mu
}\Lambda \right) ^{a}{}_{c}\left( \Lambda ^{-1}\right) ^{cb}.  \label{eq29}
\end{equation}

Requiring $D_{\mu }\left( B^{a}A_{a}\right) =\partial _{\mu }\left(
B^{a}A_{a}\right) $, since $B^{a}A_{a}$ is a scalar, one gets 
\begin{equation}
D_{\mu }B_{a}=\partial _{\mu }B_{a}-\omega _{\mu }{}^{b}{}_{a}B_{b}.
\label{eq30}
\end{equation}

The covariant derivative $\nabla _{\mu }$ of an object $B_{\nu }$ defined in
the Riemannian ma\-ni\-fold is given, as usual, as 
\begin{equation}
\nabla _{\mu }B_{\nu }=\partial _{\mu }B_{\nu }-\Gamma _{\mu \nu }{}^{\alpha
}B_{\alpha },  \label{eq30b}
\end{equation}
where $\Gamma _{\mu \nu }{}^{\alpha }$ is the connection on the ${\cal R}%
^{4} $ manifold. Then the total covariant derivative $\nabla _{\mu }$ of a
quantity $B_{\nu }{}^{a}$, with Lorentzian and Riemannian indexes, will be
given by 
\begin{equation}
\nabla _{\mu }B_{\nu }{}^{a}=D_{\mu }B_{\nu }{}^{a}-\Gamma _{\mu \nu
}{}^{\alpha }B_{\alpha }{}^{a},  \label{eq31}
\end{equation}
or 
\begin{equation}
\nabla _{\mu }B^{\nu a}=D_{\mu }B^{\nu a}+\Gamma _{\mu \alpha }{}^{\nu
}B^{\alpha a}.  \label{eq32}
\end{equation}

The relation between connections $\omega _{\mu }{}^{ab}$ and $\Gamma _{\mu
\nu }{}^{\alpha }$ can be found by requirement that $\nabla _{\mu }e_{\nu
}{}^{a}=0$, which implies that 
\begin{equation}
\omega _{\mu }{}^{ab}=e_{\alpha }{}^{a}e^{\nu b}\Gamma _{\mu \nu }{}^{\alpha
}-e^{\nu b}\partial _{\mu }e_{\nu }{}^{a}.  \label{eq33}
\end{equation}
Moreover, the metricity condition $\nabla _{\mu }g^{\nu \alpha }=0$ will
imply that $\omega _{\mu }{}^{ab}=-\omega _{\mu }{}^{ba}$. Besides this, it
is easy to see that $\nabla _{\mu }A^{\mu }=\nabla _{a}A^{a}$, where $\nabla
_{a}=e^{\mu }{}_{a}\nabla _{\mu }$, and that $e^{\nu }{}_{a}D_{\mu
}B^{a}=\nabla _{\mu }B^{\nu }=\nabla _{\mu }\left( e^{\nu
}{}_{a}B^{a}\right) =e^{\nu }{}_{a}\nabla _{\mu }B^{a}$.

\subsection{Generalized DKP equation}

The procedure to generalize DKP\ equation is very similar to the case of
Dirac one \cite{Sabbata}. First we will generalize the matrices $\beta ^{a}$
defined on flat Minkowski manifold ${\cal M}^{4}$ obeying equation (\ref{eq2}%
) to matrices $\beta ^{\mu }$ defined on Riemannian manifold ${\cal R}^{4}$
by 
\begin{equation}
\beta ^{\mu }=e^{\mu }{}_{a}\beta ^{a},  \label{eq34}
\end{equation}
that will satisfy 
\begin{equation}
\beta ^{\mu }\beta ^{\nu }\beta ^{\alpha }+\beta ^{\alpha }\beta ^{\nu
}\beta ^{\mu }=\beta ^{\mu }g^{\nu \alpha }+\beta ^{\alpha }g^{\nu \mu },
\label{eq35}
\end{equation}
as it can be shown using the properties of $\beta ^{a}$ and equation (\ref
{eq24}).

Moreover, under a local infinitesimal Lorentz transformation on ${\cal M}%
^{4} $, the multicomponent field $\psi $ will transform as given by
equations (\ref{eq7}) and (\ref{eq9}) so that its variation will be 
\begin{equation}
\delta \psi =\frac{1}{2}\omega _{ab}S^{ab}\psi  \label{eq36}
\end{equation}
and the variation of the tetrad vectors will be 
\begin{equation}
\delta e_{\mu }{}^{a}=\omega ^{a}{}_{b}e_{\mu }{}^{b}.  \label{eq37}
\end{equation}

If we consider two nearby points $x_{1}$ and $x_{2}$ with the local tetrads $%
e_{\mu }{}^{a}\left( x_{1}\right) $ and $e_{\mu }{}^{a}\left( x_{2}\right) $
then $\psi \left( x_{1}\right) $ and $\psi \left( x_{2}\right) $ are the
field $\psi $ referred to these tetrads, respectively. Performing a parallel
displacement \ from $x_{1}$ to $x_{2}$ on the tetrad $e_{\mu }{}^{a}\left(
x_{1}\right) $ we get a new one denoted by $e%
{\acute{}}%
_{\mu }{}^{a}\left( x_{2}\right) $ and the field $\psi $ at $x_{2}$ with
respect to this new tetrad will be $\psi 
{\acute{}}%
\left( x_{2}\right) $. Then, the covariant differential of the field can be
defined as 
\begin{equation}
D\psi =dx^{a}\nabla _{a}\psi =dx^{\mu }\nabla _{\mu }\psi =\psi ^{\prime
}\left( x_{2}\right) -\psi \left( x_{1}\right) ,  \label{eq38}
\end{equation}
or 
\begin{equation}
D\psi =\psi \left( x_{2}\right) -\psi \left( x_{1}\right) -\left[ \psi
\left( x_{2}\right) -\psi ^{\prime }\left( x_{2}\right) \right] ,
\label{eq39}
\end{equation}
where we separated the translation term $\psi \left( x_{2}\right) -\psi
\left( x_{1}\right) $ from the local Lorentz transformation term $\psi 
{\acute{}}%
\left( x_{2}\right) -\psi \left( x_{2}\right) $. Explicitly we have 
\begin{equation}
\psi \left( x_{2}\right) -\psi \left( x_{1}\right) =dx^{a}\partial _{a}\psi
=dx^{\mu }\partial _{\mu }\psi ,  \label{eq40}
\end{equation}
and 
\begin{equation}
\psi \left( x_{2}\right) -\psi ^{\prime }\left( x_{2}\right) =-\frac{1}{2}%
\omega _{ab}S^{ab},  \label{eq41}
\end{equation}
so that 
\begin{equation}
D\psi =dx^{a}\partial _{a}\psi +\frac{1}{2}\omega _{ab}S^{ab}\psi .
\label{eq42}
\end{equation}

From the expression (\ref{eq30}) for the Lorentz covariant derivative we see
that the variation of the tetrad vector $e_{\mu }{}^{a}$ under Lorentz
transformations on the ${\cal M}^{4}$ manifold is 
\begin{equation}
\delta e_{\mu }{}^{a}=\omega _{\nu }{}^{a}{}_{b}e_{\mu }{}^{b}dx^{\nu },
\label{eq43}
\end{equation}
so, comparing with equation (\ref{eq37}) we can identify 
\begin{equation}
\omega ^{ab}=\omega _{\nu }{}^{ab}dx^{\nu }  \label{eq44}
\end{equation}
and rewrite (\ref{eq42}) as 
\begin{equation}
D\psi =dx^{\mu }\left( \partial _{\mu }+\frac{1}{2}\omega _{\mu
ab}S^{ab}\right) \psi .  \label{eq45}
\end{equation}
From this expression we can obtain the Lorentz covariant derivative $D_{\mu
} $ of field $\psi $ and, since $\psi $ has no Riemannian index, this
derivative will be equal to the total covariant derivative $\nabla _{\mu }$
of $\psi $; so we have that 
\begin{equation}
\nabla _{\mu }\psi =D_{\mu }\psi =\left( \partial _{\mu }+\frac{1}{2}\omega
_{\mu ab}S^{ab}\right) \psi .  \label{eq46}
\end{equation}

Applying the hermitian conjugation to this expression and using the
definition of $\overline{\psi }$ we get 
\begin{equation}
\nabla _{\mu }\overline{\psi }=\partial _{\mu }\overline{\psi }-\frac{1}{2}%
\omega _{\mu ab}\overline{\psi }S^{ab}.  \label{eq47}
\end{equation}

Now we can write the generalized expression for the $\psi $ field Lagrangian 
\begin{equation}
{\cal L}=\sqrt{-g}\left[ \frac{i}{2}\left( \overline{\psi }\beta ^{\mu
}\nabla _{\mu }\psi -\nabla _{\mu }\overline{\psi }\beta ^{\mu }\psi \right)
-m\overline{\psi }\psi \right] ,  \label{eq48}
\end{equation}
that can be written explicitly as

\begin{eqnarray}
{\cal L} &=&\sqrt{-g}\left[ \frac{i}{2}\left( \overline{\psi }\beta ^{\mu
}\left( \partial _{\mu }\psi +\frac{1}{2}\omega _{\mu ab}S^{ab}\psi \right)
\right. \right. .  \nonumber \\
&&\left. \left. -\left( \partial _{\mu }\overline{\psi }-\frac{1}{2}\omega
_{\mu ab}\overline{\psi }S^{ab}\right) \beta ^{\mu }\psi \right) -m\overline{%
\psi }\psi \right] .  \label{eq49}
\end{eqnarray}

So we have 
\begin{equation}
\frac{\partial {\cal L}}{\partial \overline{\psi }}=\sqrt{-g}\left[ \frac{i}{%
2}\left( \beta ^{\mu }\partial _{\mu }\psi +\omega _{\mu ab}\beta ^{\mu
}S^{ab}\psi -\omega _{\mu }{}^{\mu }{}_{b}\beta ^{b}\psi \right) -m\psi %
\right] ,  \label{eq50}
\end{equation}
and 
\begin{equation}
\partial _{\nu }\frac{\partial {\cal L}}{\partial \left( \partial _{\nu }%
\overline{\psi }\right) }=-\frac{i}{2}\left[ \partial _{\nu }\left( \sqrt{-g}%
\beta ^{\nu }\right) \psi +\sqrt{-g}\beta ^{\nu }\partial _{\nu }\psi \right]
,  \label{eq51}
\end{equation}
\begin{equation}
\partial _{\nu }\frac{\partial {\cal L}}{\partial \left( \partial _{\nu }%
\overline{\psi }\right) }=-\frac{i}{2}\left[ \sqrt{-g}\omega _{\mu }{}^{\mu
}{}_{a}\beta ^{a}\psi +\sqrt{-g}\beta ^{\nu }\partial _{\nu }\psi \right] ,
\label{eq52}
\end{equation}
where we used the result \cite{Sabbata} 
\begin{equation}
\partial _{\nu }\left( \sqrt{-g}e^{\nu }{}_{a}\right) =\sqrt{-g}\omega
_{b}{}^{b}{}_{a}=\sqrt{-g}\omega _{\mu }{}^{\mu }{}_{a}  \label{eq53}
\end{equation}
valid for a Riemannian manifold.

Combining these results we get the generalized DKP equation of motion for
Riemannian space-times 
\begin{equation}
i\beta ^{\mu }\nabla _{\mu }\psi -m\psi =0.  \label{eq54}
\end{equation}

\section{The equivalence with KG and Proca equations}

Now we will analyze the equivalence between DKP and KG and Proca theories in
Riemannian space-times. In order to do this we will generalize the operators
defined in Section 2 to select the spin 0 and 1 sectors and apply them to
the generalized DKP equation. Then we can compare the obtained results with
KG and Proca equations in Riemannian space-times, in a way analogous to what
was done in the Minkowskian free field case.

\subsection{The spin 0 case: Equivalence with KG}

From the $P^{a}$ operators defined in Section 2 we can construct the
generalized projectors $P^{\mu }$ as 
\begin{equation}
P^{\mu }=e^{\mu }{}_{a}P^{a}=e^{\mu }{}_{a}P\beta ^{a}=P\beta ^{\mu },
\label{eq55}
\end{equation}
where $P$ is given by equation (\ref{eq11}) in terms of the matrices $\beta
^{a}$. Using equations (\ref{eq13}) and (\ref{eq24}) it is easy to verify
that 
\begin{equation}
P^{\mu }\beta ^{\nu }=Pg^{\mu \nu },\ PS^{\mu }{}^{\nu }=S^{\mu }{}^{\nu
}P=0,\ P^{\mu }S^{\alpha \nu }=\left( g^{\mu \alpha }P^{\nu }-g^{\mu \nu
}P^{\alpha }\right) .  \label{eq56}
\end{equation}

As each component of $P\psi $ was shown to be a scalar under Lorentz
transformation on the Min\-kows\-ki tangent manifold ${\cal M}^{4}$ they
will also be scalars under general coordinate transformations on the
Riemannian manifold ${\cal R}^{4}$. Similarly, each component of $P^{a}\psi $
was shown to be a vector under Lorentz transformations so that the
components of $P^{\mu }\psi $ will also be vectors under general coordinate
transformations \cite{Sabbata}. So, one should expect that the total
covariant derivative $\nabla _{\mu }$ of $P\psi $ and $P^{\mu }\psi $ should
be that of a scalar and a vector, respectively. As a matter of fact, we have
that 
\begin{equation}
\nabla _{\mu }\left( P\psi \right) =D_{\mu }\left( P\psi \right) =\partial
_{\mu }\left( P\psi \right) +\frac{1}{2}\omega _{\mu ab}S^{ab}P\psi
=\partial _{\mu }\left( P\psi \right)  \label{eq56b}
\end{equation}
and 
\begin{eqnarray}
\nabla _{\mu }\left( P^{\nu }\psi \right) &=&e^{\nu }{}_{c}\nabla _{\mu
}\left( P^{c}\psi \right) =e^{\nu }{}_{c}D_{\mu }\left( P^{c}\psi \right) 
\nonumber \\
&=&e^{\nu }{}_{c}\left[ \partial _{\mu }\left( P^{c}\psi \right) +\frac{1}{2}%
\omega _{\mu ab}S^{ab}P^{c}\psi +\omega _{\mu }{}^{c}{}_{b}P^{b}\psi \right]
,  \label{eq56c}
\end{eqnarray}
\begin{equation}
\nabla _{\mu }\left( P^{\nu }\psi \right) =e^{\nu }{}_{c}\left[ \partial
_{\mu }\left( P^{c}\psi \right) +\omega _{\mu }{}^{c}{}_{b}P^{b}\psi \right]
,  \label{eq56d}
\end{equation}
since $S^{ab}P^{c}=S^{ab}P\beta ^{c}=0$, so that the use of equation (\ref
{eq33}) results in 
\begin{equation}
\nabla _{\mu }\left( P^{\nu }\psi \right) =\partial _{\mu }\left( P^{\nu
}\psi \right) +\Gamma _{\mu \beta }{}^{\nu }P^{\beta }\psi .  \label{eq56e}
\end{equation}
These results show that we can calculate the total covariant derivatives $%
\nabla _{\mu }$ of $P\psi $ and $P^{\mu }\psi $ by applying the derivative
to each of their components as if they were, respectively, a scalar and a
vector on the Riemannian manifold ${\cal R}^{4}$ and neglect its matrix
character. Similarly, $P\psi $ and $P^{a}\psi $ can also be treated as
scalar and vector, respectively, when calculating the Lorentz covariant
derivative $D_{\mu }$.

Moreover, we can also see that 
\begin{equation}
P\nabla _{\mu }\psi =\partial _{\mu }\left( P\psi \right) +\frac{1}{2}\omega
_{\mu ab}PS^{ab}\psi =\partial _{\mu }\left( P\psi \right)  \label{eq56f}
\end{equation}
and 
\begin{equation}
P^{\nu }\nabla _{\mu }\psi =e^{\nu }{}_{c}P^{c}\nabla _{\mu }\psi =e^{\nu
}{}_{c}\left[ \partial _{\mu }\left( P^{c}\psi \right) +\frac{1}{2}\omega
_{\mu ab}P^{c}S^{ab}\psi \right] ,  \label{eq56g}
\end{equation}
\begin{equation}
P^{\nu }\nabla _{\mu }\psi =e^{\nu }{}_{c}\left[ \partial _{\mu }\left(
P^{c}\psi \right) +\omega _{\mu }{}^{c}{}_{b}P^{b}\psi \right] =\nabla _{\mu
}\left( P^{\nu }\psi \right) ,  \label{eq56h}
\end{equation}
so that it becomes obvious that $P\nabla _{\mu }\psi =\nabla _{\mu }\left(
P\psi \right) $ and $P^{\nu }\nabla _{\mu }\psi =\nabla _{\mu }\left( P^{\nu
}\psi \right) $.

Now, applying the operators $P^{\mu }$ and $P$ to the generalized DKP
equation (\ref{eq54}), we get 
\begin{equation}
P^{\mu }\psi =\frac{i}{m}\nabla ^{\mu }\left( P\psi \right) =\frac{i}{m}%
\partial ^{\mu }\left( P\psi \right) ,  \label{eq57}
\end{equation}
and 
\begin{equation}
\nabla _{\mu }\left( P^{\mu }\psi \right) =\frac{m}{i}\left( P\psi \right) .
\label{eq58}
\end{equation}

Combining equations (\ref{eq57}) and (\ref{eq58}) we obtain the generalized
KG equation 
\begin{equation}
\nabla _{\mu }\nabla ^{\mu }\left( P\psi \right) +m^{2}\left( P\psi \right)
=\nabla _{a}\nabla ^{a}\left( P\psi \right) +m^{2}\left( P\psi \right) =0.
\label{eq59}
\end{equation}

These results make clear that when we select the spin 0 sector of the
generalized DKP equation (\ref{eq54}), describing a scalar particle on a
Riemannian manifold, we get a complete equivalence with the generalized KG
equation.

Once more we can make this equivalence more explicit using the specific
choice of the matrices $\beta ^{a}$ that satisfies condition (\ref{eq19}).
Then we get as result 
\begin{equation}
P\psi =\left( 
\begin{array}{c}
0_{4\times 1} \\ 
\psi _{4}
\end{array}
\right) ,\ P^{\mu }\psi =e^{\mu a}P_{a}\psi =e^{\mu a}{}\left( 
\begin{array}{c}
0_{4\times 1} \\ 
\psi _{a}
\end{array}
\right) ,  \label{eq60}
\end{equation}
\begin{equation}
\psi =\left( 
\begin{array}{c}
\frac{i}{\sqrt{m}}\nabla _{a}\varphi \\ 
\sqrt{m}\varphi
\end{array}
\right) ,\ \nabla _{\mu }\nabla ^{\mu }\varphi +m^{2}\varphi =0,
\label{eq61}
\end{equation}
and, finally, we note that it is possible to make a change in the
representation of $\beta $ matrices in such a way that the form of DKP field 
$\psi $ becomes 
\begin{equation}
\psi \rightarrow \psi _{R}=\left( 
\begin{array}{c}
\frac{i}{\sqrt{m}}\nabla _{\mu }\varphi \\ 
\sqrt{m}\varphi
\end{array}
\right) .  \label{eq62}
\end{equation}

\subsection{The spin 1 case: Equivalence with Proca}

Similarly to the spin 0 case we can also generalize the operators $R^{a}$
and $R^{ab}$ defined in Section 2 as 
\begin{equation}
R^{\mu }=e^{\mu }{}_{c}R^{c},\ R^{\mu \nu }=e^{\mu }{}_{c}e^{\nu
}{}_{d}R^{cd},  \label{eq63}
\end{equation}
which can be seen to satisfy the relations 
\begin{equation}
R^{\mu \nu }=-R^{\nu \mu },  \label{eq64}
\end{equation}
\begin{equation}
R^{\mu }\beta ^{\nu }\beta ^{\rho }=R^{\mu \nu }\beta ^{\rho }=g^{\nu \rho
}R^{\mu }-g^{\mu \rho }R^{\nu },  \label{eq65}
\end{equation}
\begin{equation}
R^{\mu }S^{\nu \rho }=g^{\mu \nu }R^{\rho }-g^{\mu \rho }R^{\nu },\ S^{\nu
\rho }R^{\mu }=0,  \label{eq66}
\end{equation}
and 
\begin{equation}
R^{\mu \nu }S^{\alpha \beta }=g^{\nu \alpha }R^{\mu \beta }-g^{\mu \alpha
}R^{\nu \beta }-g^{\nu \beta }R^{\mu \alpha }+g^{\mu \beta }R^{\nu \alpha }.
\label{eq67}
\end{equation}

Analogously to the case of $P\psi $ and $P^{\mu }\psi $, we would expect the
total covariant derivatives $\nabla _{\mu }$ of $R^{\mu }\psi $ and $R^{\mu
\nu }\psi $ to be, respectively, those of a vector and a tensor since $%
R^{a}\psi $ is a vector and $R^{ab}\psi $ a tensor on the Min\-kows\-ki
tangent manifold ${\cal M}^{4}$. This turns out to be true since 
\begin{equation}
\nabla _{\mu }\left( R^{\nu }\psi \right) =e^{\nu }{}_{c}\nabla _{\mu
}\left( R^{c}\psi \right) =e^{\nu }{}_{c}\left[ \partial _{\mu }\left(
R^{c}\psi \right) +\frac{1}{2}\omega _{\mu ab}S^{ab}R^{c}\psi +\omega _{\mu
}{}^{c}{}_{b}R^{b}\psi \right] ,  \label{eq68}
\end{equation}
\begin{equation}
\nabla _{\mu }\left( R^{\nu }\psi \right) =e^{\nu }{}_{c}\left[ \partial
_{\mu }\left( R^{c}\psi \right) +\omega _{\mu }{}^{c}{}_{b}R^{b}\psi \right]
,  \label{eq69}
\end{equation}
and 
\begin{equation}
\nabla _{\mu }\left( R^{\alpha \nu }\psi \right) =e^{\alpha }{}_{c}e^{\nu
}{}_{d}\nabla _{\mu }\left( R^{cd}\psi \right) ,  \label{eq70}
\end{equation}
\begin{equation}
\nabla _{\mu }\left( R^{\alpha \nu }\psi \right) =e^{\alpha }{}_{c}e^{\nu
}{}_{d}\left[ \partial _{\mu }\left( R^{cd}\psi \right) +\frac{1}{2}\omega
_{\mu ab}S^{ab}R^{cd}\psi +\omega _{\mu }{}^{c}{}_{b}R^{bd}\psi +\omega
_{\mu }{}^{d}{}_{b}R^{cb}\psi \right] ,  \label{eq71}
\end{equation}
\begin{equation}
\nabla _{\mu }\left( R^{\alpha \nu }\psi \right) =e^{\alpha }{}_{c}e^{\nu
}{}_{d}\left[ \partial _{\mu }\left( R^{cd}\psi \right) +\omega _{\mu
}{}^{c}{}_{b}R^{bd}\psi +\omega _{\mu }{}^{d}{}_{b}R^{cb}\psi \right] ,
\label{eq72}
\end{equation}
where we used $S^{ab}R^{cd}=S^{ab}R^{c}\beta ^{d}$ and $S^{ab}R^{c}=0$, so
that using equation (\ref{eq33}) in equations (\ref{eq69}) and (\ref{eq72})
results in 
\begin{equation}
\nabla _{\mu }\left( R^{\nu }\psi \right) =e^{\nu }{}_{c}\left[ \partial
_{\mu }\left( R^{c}\psi \right) +\Gamma _{\mu \beta }{}^{\nu }R^{\beta }\psi %
\right] ,  \label{eq73}
\end{equation}
and 
\begin{equation}
\nabla _{\mu }\left( R^{\alpha \nu }\psi \right) =e^{\alpha }{}_{c}e^{\nu
}{}_{d}\left[ \partial _{\mu }\left( R^{cd}\psi \right) +\Gamma _{\mu \beta
}{}^{\alpha }R^{\beta \nu }\psi +\Gamma _{\mu \beta }{}^{\nu }R^{\alpha
\beta }\psi \right] .  \label{eq74}
\end{equation}
So we can neglect the matrix character of $R^{\mu }\psi $ and $R^{\mu \nu
}\psi $ when calculating their total covariant derivative and proceed by
applying the derivative to each of \ their components as if they were usual
vector and tensor, respectively, on the Riemannian manifold ${\cal R}^{4}$.
Moreover we can also show that 
\begin{equation}
R^{\nu }\nabla _{\mu }\psi =e^{\nu }{}_{c}R^{c}\nabla _{\mu }\psi =e^{\nu
}{}_{c}R^{c}\left[ \partial _{\mu }\left( \psi \right) +\frac{1}{2}\omega
_{\mu ab}S^{ab}\psi \right] ,  \label{eq75}
\end{equation}
\begin{equation}
R^{\nu }\nabla _{\mu }\psi =e^{\nu }{}_{c}\left[ \partial _{\mu }\left(
R^{c}\psi \right) +\frac{1}{2}\omega _{\mu ab}R^{c}S^{ab}\psi \right] ,
\label{eq76}
\end{equation}
\begin{equation}
R^{\nu }\nabla _{\mu }\psi =e^{\nu }{}_{c}\left[ \partial _{\mu }\left(
R^{c}\psi \right) +\omega _{\mu }{}^{c}{}_{b}R^{b}\psi \right] ,
\label{eq77}
\end{equation}
and 
\begin{equation}
R^{\alpha \nu }\nabla _{\mu }\psi =e^{\alpha }{}_{c}e^{\nu
}{}_{d}R^{cd}\nabla _{\mu }\psi =e^{\alpha }{}_{c}e^{\nu }{}_{d}\left[
\partial _{\mu }\left( R^{cd}\psi \right) +\frac{1}{2}\omega _{\mu
ab}R^{cd}S^{ab}\psi \right] ,  \label{eq78}
\end{equation}
\begin{equation}
R^{\alpha \nu }\nabla _{\mu }\psi =e^{\alpha }{}_{c}e^{\nu }{}_{d}\left[
\partial _{\mu }\left( R^{cd}\psi \right) +\omega _{\mu
}{}^{d}{}_{b}R^{cb}\psi +\omega _{\mu }{}^{c}{}_{b}R^{bd}\psi \right] ,
\label{eq79}
\end{equation}
so that it becomes clear that $R^{\nu }\nabla _{\mu }\psi =\nabla _{\mu
}\left( R^{\nu }\psi \right) $ and $R^{\alpha \nu }\nabla _{\mu }\psi
=\nabla _{\mu }\left( R^{\alpha \nu }\psi \right) $.

Now we can use the operators $R^{\mu }$ and $R^{\mu \nu }$ on the
generalized DKP equation (\ref{eq54}), obtaining 
\begin{equation}
\nabla _{\lambda }\left( R^{\mu \lambda }\psi \right) =\frac{m}{i}R^{\mu
}\psi  \label{eq80}
\end{equation}
and 
\begin{equation}
R^{\mu \nu }\psi =-\frac{i}{m}U^{\mu \nu },  \label{eq81}
\end{equation}
where now we have a covariant strength tensor 
\begin{equation}
U^{\mu \nu }=\nabla ^{\mu }R^{\nu }\psi -\nabla ^{\nu }R^{\mu }\psi .
\label{eq82}
\end{equation}

Combining equations (\ref{eq80}) and (\ref{eq81}) we get the generalized
Proca equation 
\begin{equation}
\nabla _{\lambda }\left( U^{\lambda \mu }\right) +m^{2}R^{\mu }\psi =0.
\label{eq83}
\end{equation}

This makes clear that the spin 1 sector of the generalized DKP equation (\ref
{eq54}), describing a massive vectorial particle on a Riemannian manifold,
is completely equivalent to the generalized Proca equation.

\section{Conclusions and comments}

In this work we analyzed the generalization of DKP theory to Riemannian
space-times. We followed the standard procedure to perform the
generalization by constructing the Riemmanian quantities from the flat
space-time ones through the use of tetrad formalism. We have obtained a
first order relativistic wave equation that describes spin 0 and 1 particles
coupled to the gravitational field. Then we used the operators that select
the spin 0 and 1 sectors of the theory and obtained, respectively, the
generalized KG and Proca equations for spin 0 and 1 bosons in Riemannian
space-times, proving the equivalence between the theories. It should be also
mentioned that the form (\ref{eq61}) for DKP field when the matrices $\beta $
satisfy condition (\ref{eq60}) is, in principle, analogous to the result
obtained when we consider the interaction with eletromagnetic field: the
``gradient'' components of free field case (the first four components) are
replaced by the covariant derivatives \cite{Pimentel 2}.

This analogy is not clearly manifested in the case of the spin 0 sector
because the covariant derivatives in a curved space-time will, of course,
simply reduce to the usual derivatives due to the scalar character of $%
\varphi $. But if we perform for the spin 1 sector of DKP field in
Riemannian space-times the same construction made in equations (\ref{eq20-16}%
) to (\ref{eq20-22}) in Minkowski space-time through a specific choice of a
10 degree representation of the $\beta $ matrices, we will find that the
usual derivatives will be replaced by covariant ones in those equations,
changing the forms of the components $\psi _{4}$ to $\psi _{9}$ of $\psi $
field. This procedure shows, together with the result of \cite{Pimentel 2},
that care must be taken in account when using specific representations of
the $\beta $ matrices to obtain the components of DKP field $\psi $: the
forms of the components obtained in the free field case may not be usable to
interacting fields, requiring a new construction.

The perspectives for future developments are diverse and some of them are
currently under our study. We could mention the application of the DKP\
theory to the electromagnetic field in Riemannian space-times. Moreover, it
is interesting to look for the construction of a generalization similar to
the one proposed by Dirac to its electron equation \cite{Dirac}.
Subsequently, it comes the generalizations to Riemann-Cartan space-times 
\cite{Sabbata} and to teleparallel descripition of gravity \cite{Hayashi,
Andrade, Maluf} as a step to compare the question of coupling of DKP fields
to torsion in both theories, similarly to what has been done to usual spin 0
and 1 formalisms \cite{Andrade 1, Andrade 2}. Independently, it will be
analyzed the quantic processes in DKP\ theory with gravitation viewed as an
external field, as it has already been done with DKP field in an external
electromagnetic field \cite{Pimentel 1}.

\section{Acknowlegdments}

J.T.L. and B.M.P. would like to thank CAPES's PICDT program and CNPq,
respectively, for partial support. R.G.T. thanks CAPES\ for full support.
The authors also wish to thank Professor V. Ya. Fainberg by his critical
reading of our manuscript.

\end{document}